\documentstyle[12pt,epsf]{article}
\input{psfig.sty}

\title{Are spin junction transistors suitable for signal processing?}
\author{S. Bandyopadhyay$\thanks{Corresponding author. E-mail: sbandy@vcu.edu}$
\\Department of Electrical and Computer Engineering\\ Virginia 
Commonwealth University, Richmond, VA 23284, USA\\
\\M. Cahay\\
Department of Electrical and Computer Engineering and Computer Science \\
University of Cincinnati, Cincinnati, OH 45221, USA}

\begin{document}

\maketitle

\begin{abstract}

A number of spintronic junction transistors, that exploit the spin degree of 
freedom of an electron in addition to the charge degree of freedom, have been 
proposed to provide 
simultaneous  non-volatile storage and signal processing  functionality. Here, 
we show that some of these transistors unfortunately may not have sufficient 
voltage and current gains for 
signal processing. This is primarily because of a large output ac conductance 
and 
poor isolation between input and output.  The latter also hinders 
unidirectional propagation of logic signal from the input of a logic gate to the 
output. Other versions of these transistors appear to have better gain and 
isolation, but not better than those of a conventional transistor. Therefore, 
these devices may not {\it improve} state-of-the-art signal processing 
capability, although they may provide additional functionality by offering  
non-volatile storage. They may also have niche applications in non-linear 
circuits.
\end{abstract}


\pagebreak

A number of spin analogs of conventional 
transistors have been proposed recently with a view to providing both 
signal processing and {\it non-volatile} storage functions 
with the same device.  In this letter, we examine whether these devices meet the 
stringent requirements for signal processing. 

``Analog'' signal processing usually requires devices to have both large current 
 and voltage gain for power amplification. ``Digital'' logic devices, on the 
other hand, have other critical requirements \cite{hodges}, three of which are 
that (i) the 
device must have a large voltage gain  to regenerate logic levels at signal 
nodes, (ii) a large current gain for 
adequate fan out, and (iii) no feedback from the output terminal to the input 
terminal (a property known as ``isolation between input and output'') so that 
logic signal can
propagate {\it unidirectionally} from the input to the output. A conventional 
transistor 
has all these attributes and therefore has become the workhorse of analog and 
digital (as well as ``mixed signal'') circuits. Spin transistors need to  have 
the same attributes to be useful.

There are two basic types of transistors: the field effect type (FET) and the 
bipolar junction type (BJT). In this  letter, we will focus on spin analogs of 
the latter, since we had already examined the device potentials 
of spintronic FETs earlier \cite{apl}. Two varieties of 
spin-BJTs have been proposed: (i) unipolar spin junction 
transistor (USJT) that mimics a conventional BJT \cite{flatte}, and (ii) bipolar 
spin junction transistor (BSJT) whose only difference with a conventional BJT is 
that the base is ferromagnetic and has a non-zero equilibrium spin polarization 
\cite{fabian,flatte1,fabian2}. 

We consider first the USJT of ref. \cite{flatte}. This device consists of three 
layers of spin polarized material which act as emitter, base and collector. In 
the emitter and collector layers, spin of one kind (say, ``downspin'') is 
majority, while in the base, spin of the other kind (``upspin'') is majority. 
Since all three layers can have the same charge polarity,  the 
device is ``unipolar '', but operationally it mimics a conventional BJT. Ref. 
\cite{flatte} derived the expressions for the collector 
current $I_C$ and emitter current $I_E$ as functions  of the emitter-base bias 
$V_{EB}$ and collector-base bias $V_{CB}$:
\begin{equation}
I_C = - {{q J_0}\over{sinh(W/L)}} \left [ \left (e^{-q V_{EB}/kT} - 1 \right ) - 
\left (e^{-q V_{CB}/kT} - 1 \right ) cosh(W/L) \right ]  - qJ_0 \left [ e^{q 
V_{CB}/kT} - 1 \right ]
\end{equation}
\begin{equation}
I_E  =  - {{q J_0}\over{sinh(W/L)}} \left [ \left (e^{-q V_{EB}/kT} - 1 \right ) 
cosh(W/L) - \left (e^{-q V_{CB}/kT} - 1 \right )  \right ]   + qJ_0 \left [ e^{q 
V_{EB}/kT} - 1 \right ]
\end{equation}
where $qJ_0$ is the constant saturation current, $W$ is the base width and $L$ 
is the minority spin diffusion length (assumed  same in all layers) 
\cite{flatte2}. In the 
active mode of operation, $V_{EB}$ $<$ 0 and $V_{CB}$ $ > $ 0.

\begin{figure}
\centerline{\psfig{figure=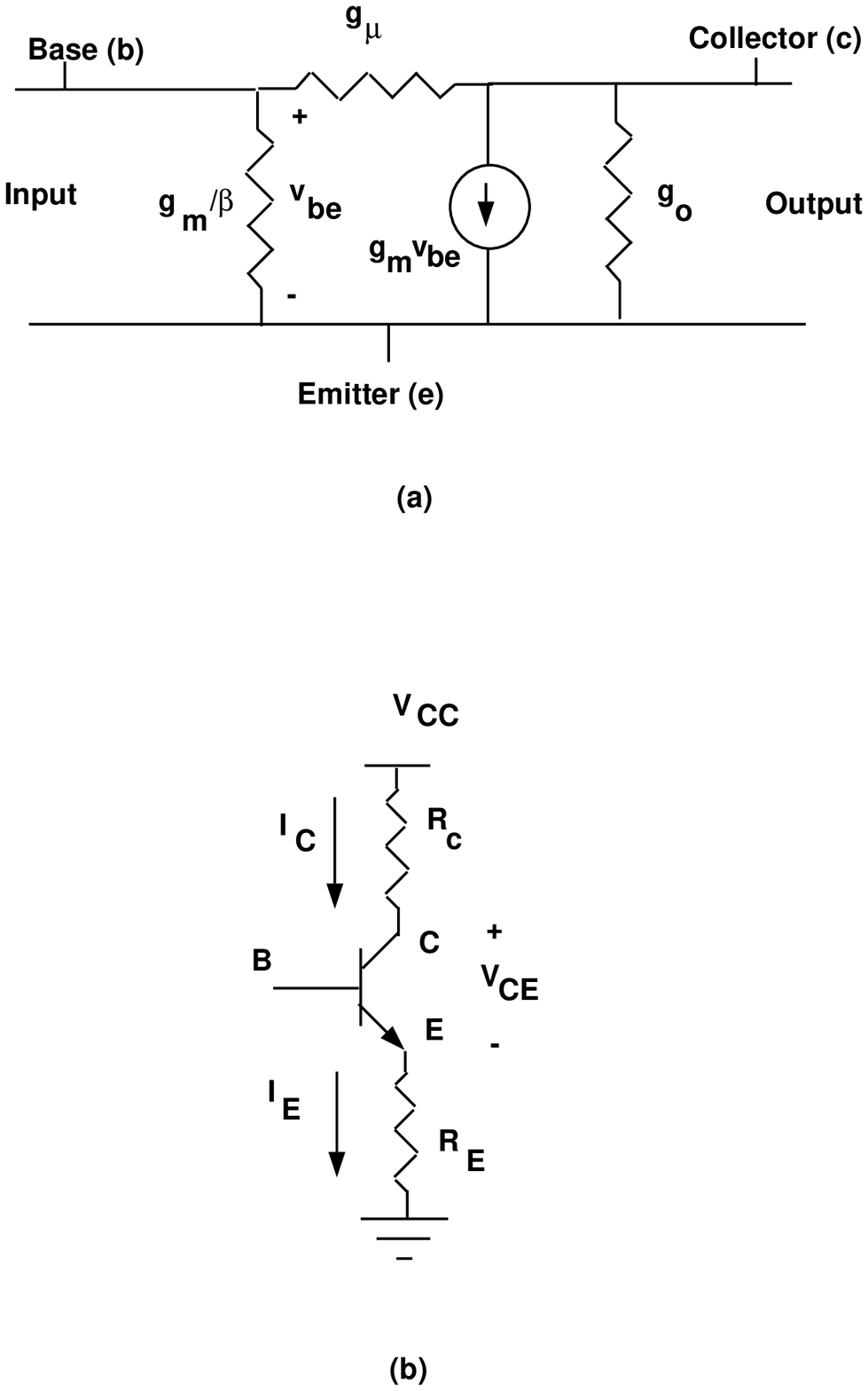,height=5.5in,width=4.5in}}
\caption{(a) Universal low-frequency small-signal equivalent circuit of a BJT 
which 
applies to both USJT and BSJT. The small signal conductances are shown. (b) A 
biased transistor. }
\end{figure}

In Fig. 1(a), we show the universal low-frequency small-signal equivalent 
circuit 
of a BJT \cite{sedra}, which applies to both the USJT and BSJT.
First, we will derive expressions for two small-signal parameters -- the 
transconductance $g_m$ and the output conductance $g_o$ -- in the ``active 
mode''.  They are given by \cite{pierret}:
\begin{eqnarray}
g_m & = & {{\partial I_C}\over{\partial V_{EB}}}  {\Bigg\vert}_{V_{BC} = 
constant} = {{q^2 J_0}\over{kT sinh(W/L)}} e^{q|V_{EB}|/kT} \nonumber \\
g_o & = & {{\partial I_C}\over{\partial V_{CE}}}  {\Bigg\vert}_{V_{EB} = 
constant} = 
- {{\partial I_C}\over{\partial V_{CB}}}  {\Bigg\vert}_{V_{EB} = constant} = 
{{q^2 J_0}\over{kT}} \left [ e^{q|V_{CB}|/kT} - coth \left ( {{W}\over{L}} 
\right )
e^{-q|V_{CB}|/kT} \right ] \nonumber \\
&& \approx
  {{q^2 J_0}\over{kT}}e^{q|V_{CB}|/kT} ~.
  \end{eqnarray}
  
  The output conductance $g_o$ can be quite large. If we assume realistic 
values, e.g. $qJ_0$ = 1 fA, and $|V_{CB}|$ = 1 V, then $g_o$ $\approx$ 9000 S, 
which is extremely large. The situation can be improved slightly by having a  
{\it larger} spin splitting in the collector than in the emitter. In that case, 
the exponent in the last term of Equation (1) will be modified as $V_{CB} 
\rightarrow V_{CB} - \Delta_c + \Delta_b$, where $\Delta_c$ and $\Delta_b$ are 
the spin splittings in the collector and base respectively. Accordingly, $g_o$ 
$\approx$ $q^2 J_0/(kT)e^{(q|V_{CB}| - \Delta_c + \Delta_b)/kT}$.
  
  Realistically, spin splitting energies in known materials hardly exceed $kT$ 
at room temperature. Therefore, if we assume that $\Delta_c - \Delta_b$ = 25 mV, 
then $g_o$ = 3330 S, which is $\sim$ 3 times smaller, but still very large. Of 
course, we can improve the situation by choosing smaller values of $|V_{EB}|$ 
and $|V_{CB}|$, but this is not advisable. The voltage levels must be several 
times larger than the thermal voltage $kT/q$ to preserve good noise margin. 
Currently, power supply voltages are several volts and there is no movement 
towards making them millivolts.
  
  Using the above results, we find that the small signal voltage gain in the 
common-emitter or common-base configuration (the voltage gain in 
common-collector configuration is always less than unity) is \cite{sedra}
\begin{equation}
a_v = |g_m/g_o| = {{exp[(q|V_{EB}|  - q|V_{CB}| + \Delta_c - 
\Delta_b)/kT]}\over{sinh(W/L)}} ~.
\end{equation}

If we assume the previous values for the junction voltages, and $W/L$ = 0.001, 
then $a_v$ = 0.017, which is far less than unity. 

One obvious way to make the voltage gain larger than unity, is to make $|V_{EB}| 
> |V_{CB}|$. But this poses a problem. When the transistor is in the ``off'' 
state and $I_C, I_E ~\approx$ 0, the emitter-base forward bias voltage 
$|V_{EB}|$ must be approximately 0 (see Equations (1) and (2)). Therefore, if 
$|V_{EB}| > |V_{CB}|$, then $|V_{CB}|$ $\approx$ 0. Consequently, $V_{CE}$ = 
$V_{CB} ~+~ V_{BE}$ $\approx$ 0.  But now, consider the biased transistor 
circuit shown in Fig. 1(b). Kirchoff's voltage law dictates that $V_{CC}$ = $I_C 
R_C ~+~ I_E R_E ~+~ V_{CE}$. Since near the ``off'' state, $I_C$, $I_E$ and 
$V_{CE}$ are all approximately zero, then $V_{CC}$ $\approx$ 0 as well! 
Therefore, we have an inconsistency. The transistor can never be switched off 
unless the power supply voltage is vanishingly small. We can of course designate 
the ``off'' state to be one with non-zero collector and emitter currents, but 
this leads to unacceptable standby power dissipation. On the other hand, if we 
work with a vanishingly small power supply voltage, then we operate with 
unacceptable noise margin and furthermore, the logic levels (``on'' and ``off'' 
states) are not well separated, leading to unacceptable bit error rates. 
Therefore, we cannot operate under the condition that $|V_{EB}| > |V_{CB}|$. In 
other words, we are constrained to operate with a small voltage gain, which is  
good for neither digital, nor analog circuitry.

Next, we calculate the small-signal feedback conductance $g_{\mu}$ which is 
defined as $g_{\mu}$ = $\partial I_B/\partial V_{CB}$ \cite{pierret} evaluated 
at a constant value of $V_{EB}$, where $I_B$ = $I_E$ - $I_C$. As is evident from 
Fig. 1(a), the physical significance of $g_{\mu}$ is that it is a measure of 
the isolation between the output and input of the transistor since this 
conductor connects the output  to the input. Ideally, $g_{\mu}$ = 0, which 
implies perfect isolation. However, after some algebra, we can show that in the 
active mode
\begin{equation}
g_{\mu} \approx {{q^2 J_0}\over{kT}} e^{(q |V_{CB}| - \Delta_c + \Delta_b)/kT} 
\approx g_0 ~.
\end{equation}
For a normal BJT, this quantity is approximately $q^2 J_0/(kT) e^{-q 
|V_{CB}|/kT}$.
Therefore the $g_{\mu}$ for a USJT is about $e^{(2 q |V_{CB}| - \Delta_c + 
\Delta_b)/kT}$ times larger than for a normal BJT. Again, if we assume that 
$|V_{CB}|$ = 1 V and $\Delta_c - \Delta_b$ = 25 mV, we find that the $g_{\mu}$ 
for a USJT is more than 10$^{34}$ times worse than that for a conventional BJT, 
resulting in that much poorer isolation between the input and the output of the 
transistor. Isolation is an extremely important issue since there 
must exist a unilateral cause-effect relationship between the input and the 
output of a logic device. Without sufficient isolation between the input and 
output, it is impossible to ensure this relationship. To understand this issue, 
consider Fig. 1(a) if $g_{\mu}$ = 0. Then there will be no direct connection 
between the input and output. In that case, the output voltage (= -$g_m 
v_{be}/g_o$) directly depends on the input voltage $v_{be}$, but the input 
voltage $v_{be}$ is an independent variable that does not depend on the output 
voltage. Therefore, the input controls the output, and not the 
other way around. However, if $g_{\mu}$ $\neq$ 0, then there is a feedback from 
the output to the input which impairs the unilateral cause-effect relationship 
between the input and output. In fact, a non-zero $g_{\mu}$  makes the input 
voltage clearly depend on the output voltage. This does not bode well for logic 
applications.

Next, we consider the short circuit current gain of a USJT. This is given by 
\cite{sedra}
\begin{equation}
a_i = {{\beta_0 (1 - g_{\mu}/g_m)}\over{1 + \beta_0 g_\mu/g_m}}
\label{current_gain}
\end{equation}
where $\beta_0$ = $I_C/I_B$. If $g_{\mu}$ = 0, then $a_i$ = $\beta_0$, which can 
be very large. But if $g_{\mu} \neq 0$, then using Equations (3)-(6), we find 
that 
\begin{equation}
a_i \approx {{\beta_0 (1 - 1/a_v)}\over{1 + \beta_0/a_v}} ~.
\end{equation}
Since $a_v$ is small, the short circuit current gain is degraded from $\beta_0$, 
and will be small. In fact, if $a_v <<$1, then $a_i$ $\approx$ -1. This 
degradation is a consequence of a non-zero $g_{\mu}$.

We conclude that a USJT is not competitive with conventional BJTs for mainstream 
analog or digital signal processing applications since it has much 
larger output and feedback conductances. Of course, that does not preclude other 
niche applications for the USJT. 

Now, we consider the BJST device of ref. \cite{flatte2,fabian}. Since this 
device is very similar to a conventional BJT, with the only difference being 
that the base is ferromagnetic, we will be able to directly compare it with a 
conventional BJT.
 Using the current expressions derived in ref. \cite{fabian1} for an npn 
transistor, we get
\begin{eqnarray}
I_C & = & qA {{D_{nb}}\over{L_{nb}}}{{1}\over{sinh (W/L_{nb})}} n_{be} \left [ 
e^{qV_{EB}/kT} -1 \right ] \nonumber \\
&& - qA {{D_{nb}}\over{L_{nb}}}coth (W/L_{nb}) n_{bc} \left [ e^{qV_{CB}/kT} -1 
\right ] \nonumber \\
&& - qA {{D_{pc}}\over{L_{pc}}}coth (W_c/L_{pc}) p_{oc} \left [ e^{qV_{CB}/kT} 
-1 \right ] 
\end{eqnarray}

\begin{eqnarray}
I_E & = & qA {{D_{nb}}\over{L_{nb}}}coth(W/L_{nb}) n_{be} \left [ e^{qV_{EB}/kT} 
-1 \right ] \nonumber \\
&  & - qA {{D_{nb}}\over{L_{nb}}}{{1}\over{sinh (W/L_{nb})}} n_{bc}) \left [ 
e^{qV_{CB}/kT} -1 \right ] \nonumber \\
&& + qA {{D_{pe}}\over{L_{pe}}}coth (W_e/L_{pe}) p_{oe} \left [ e^{qV_{EB}/kT} 
-1 \right ] ~,
\end{eqnarray}
where $A$ is the cross-sectional area of the transistor, $W_c$ is the width of 
the collector, $W_e$ is the width of the emitter, $D_{nb}$ ($L_{nb}$) is the 
minority carrier diffusion constant (length) for electrons in the base, $D_{pc}$ 
($L_{pc}$) is the minority carrier diffusion constant (length) for holes in the 
collector,
$D_{pe}$ ($L_{pe}$) is the minority carrier diffusion constant (length) for 
holes in the emitter,  
$n_{be}$ = $(n_i^2/N_{AB})(1 + \alpha_e \alpha_{0b})/\sqrt{1 - \alpha_{0b}^2}$,
$n_{bc}$ = $(n_i^2/N_{AB})(1 + \alpha_c \alpha_{0b})/\sqrt{1 - \alpha_{0b}^2}$,
$p_{oc}$ = $(n_i^2/N_{DC})$, $p_{oe}$ = $ (n_i^2/N_{DE})$,
$n_i$ is the intrinsic carrier concentration in the material, $N_{AB}$ is the 
acceptor dopant concentration in the base, $N_{DC}$ is the donor dopant 
concentration in the collector, $N_{DE}$ is the donor dopant concentration in 
the emitter, $\alpha_e$ and $\alpha_c$ are the non-equilibrium spin 
polarizations in the emitter and collector, $\alpha_{0b}$ (= $tanh(\Delta/kT)$) 
is the equilibrium spin polarization in the base, and 2$\Delta$ is the magnitude 
of energy 
splitting between the majority and minority spin in the base.

As before, we calculate the small signal parameters in the active mode:
\begin{eqnarray} 
g_m & = & {{\partial I_C}\over{\partial V_{EB}}}  {\Bigg\vert}_{V_{BC} = 
constant} = {{q^2A}\over{kT}} {{D_{nb}}\over{L_{nb}}}{{1}\over{ sinh 
(W/L_{nb})}} {{n_i^2}\over{N_{AB}}}{{1 + \alpha_e \alpha_{0b}}\over{\sqrt{1 - 
\alpha_{0b}^2}}} e^{qV_{EB}/kT} \nonumber \\
g_o & = & {{\partial I_C}\over{\partial V_{CE}}}  {\Bigg\vert}_{V_{EB} = 
constant} = {{q^2A}\over{kT}} \left [  {{D_{nb}}\over{L_{nb}}}coth \left 
({{W}\over{L_{nb}}} \right ) {{n_i^2}\over{N_{AB}}}{{1 + \alpha_c 
\alpha_{0b}}\over{\sqrt{1 - \alpha_{0b}^2}}}  + 
 {{D_{pc}}\over{L_{pc}}}coth \left ( {{W_c}\over{L_{pc}}} \right ) 
{{n_i^2}\over{N_{DC}}} \right ] e^{qV_{CB}/kT} \nonumber \\
 g_{\mu} & = & {{\partial I_B}\over{\partial V_{CE}}} {\Bigg\vert}_{V_{EB} = 
constant} = {{q^2A}\over{kT}} \left [  {{D_{nb}}\over{L_{nb} }} {{W}\over{2 
L_{nb}}}
{{n_i^2}\over{N_{AB}}} {{1 + \alpha_c \alpha_{0b}}\over{\sqrt{1 - 
\alpha_{0b}^2}}} +  {{D_{pc}}\over{L_{pc} }}{{n_i^2}\over{N_{DC}}}coth \left( 
{{W_c}\over{L_{pc}}} \right ) \right ]
e^{qV_{CB}/kT} \nonumber \\
\end{eqnarray}

In deriving the expression for $g_{\mu}$, we have used the fact that $I_B$ = 
$I_E$ - $I_C$.

If we assume that the equilibrium spin polarization in the base is larger than 
76\% so that 
 $\Delta$ $>$ $kT$,  and additionally if we assume that the the base and 
collector dopings are about equal, as well as the base and collector widths are 
about equal, then the above expressions can be approximated as:
\begin{eqnarray}
g_m & \approx & {{q^2A}\over{2kT}} \left [ {{D_{nb}}\over{L_{nb}}}{{1}\over{ 
sinh 
(W/L_{nb})}} {{n_i^2}\over{N_{AB}}}(1 + \alpha_e tanh(\Delta/kT)) \right ] 
e^{q(V_{EB} + 
\Delta)/kT} \nonumber \\
g_o & \approx & {{q^2A}\over{2kT}} \left [  {{D_{nb}}\over{L_{nb}}}coth \left 
({{W}\over{L_{nb}}} \right ) {{n_i^2}\over{N_{AB}}} (1 + \alpha_c 
tanh(\Delta/kT)) \right ] e^{q(V_{CB} + \Delta)/kT} \nonumber \\
g_{\mu} & \approx &{{q^2A}\over{2kT}} \left [  {{D_{nb}}\over{L_{nb}}} 
{{W}\over{2 L_{nb}}} {{n_i^2}\over{N_{AB}}} (1 + \alpha_c 
tanh(\Delta/kT)) \right ] e^{q(V_{CB} + \Delta)/kT} 
\end{eqnarray}

We can now compare the parameters of a BSJT with those of a conventional BJT:

\begin{eqnarray}
{{g_m (BSJT)}\over{g_m(BJT)}} & = &  (1 + \alpha_e tanh(\Delta/kT))e^{\Delta/kT} 
\nonumber \\
{{g_o (BSJT)}\over{g_o(BJT)}} & = &  (1 + \alpha_e tanh(\Delta/kT))e^{\Delta/kT} 
\nonumber \\
{{g_{\mu} (BSJT)}\over{g_{\mu}(BJT)}} & = &  (1 + \alpha_e 
tanh(\Delta/kT))e^{\Delta/kT} \nonumber \\
{{a_v (BSJT)}\over{a_v(BJT)}} & = &  (1 + \alpha_e tanh(\Delta/kT))/(1 + 
\alpha_e tanh(\Delta/kT)) \approx 1 \nonumber \\
{{a_i (BSJT)}\over{a_i(BJT)}} & = & 1 
\end{eqnarray}
where we have used Equation (\ref{current_gain}) to evaluate the ratio of the 
current gains.

The above comparison shows that there is no significant advantage or 
disadvantage  of a BSJT compared to a conventional BJT as far as current and 
voltage gains are concerned. However, there is a drawback in terms of having a 
larger feedback conductance $g_{\mu}$ which degrades isolation between input and 
output of the device, and this impairs logic functionality. The degradation 
becomes progressively worse with increasing $\Delta$ or increasing spin 
polarization in the base. Because of this, it is unlikely that a BSJT will 
replace a BJT in digital signal processing applications.

In conclusion, we have found that spin analogs of bipolar junction transistors 
do not offer an advantage over their conventional counterparts in {\it 
mainstream} 
signal processing applications. This is consistent with our earlier finding 
regarding spin analogs of field effect transistors \cite{apl}. We stress that 
neither the proponents of Spin-BJTs, nor the proponents of Spin-FETs claimed 
explicitly that their devices have an advantage over conventional transistors in 
signal processing. The small 
signal analysis in this Letter confirms that there is indeed no such advantage. 
However,  spin transistors do have some special features that are absent in 
their conventional counterparts. They can store information via magnetism and 
perform  non-conventional tasks such as spin 
filtering \cite{flatte1}, magnetic field sensing \cite{fabian4}, etc. The 
current gains of BSJTs depend on the degree of spin polarization in the base, 
which 
can be altered with an external magnetic field using the Zeeman effect. This 
feature can be exploited to realize mixers/modulators  and other non-linear 
circuits. For example, if the base current is a sinusoid with angular frequency 
$\omega_1$ and the external magnetic field is sinusoidal with angular frequency 
$\omega_2$, then the collector current will have frequencies $\omega_1 \pm 
\omega_2$. Therefore, it appears that  the role of spin transistors is not in 
mainstream digital and analog applications, but perhaps in unusual applications 
where their unique features make them particularly suitable entities. 

The work of S. B. was supported by the Air Force Office of Scientific 
Research under grant FA9550-04-1-0261. We acknowledge fruitful discussions 
with Dr. Jaroslav Fabian and Dr. Michael Flatt\'e.

\pagebreak

\end{document}